\documentclass[prl, reprint]{revtex4-1}
\pdfoutput=1
\usepackage{amsmath}
\usepackage{amssymb}
\usepackage{graphicx}
\usepackage[utf8]{inputenc}
\usepackage{color}
\usepackage{hyperref}

\graphicspath{{figures/}}

\newcommand{\ket}[1]{|#1\rangle}
\newcommand{\bra}[1]{\langle#1|}

\begin{document}
\title{Cavity-controlled ultracold chemistry}
\author{Tobias Kampschulte}\email{tobias.kampschulte@uni-ulm.de}
\author{Johannes Hecker Denschlag}
\affiliation{Institut für Quantenmaterie and Center for Integrated Quantum Science and Technology (IQ$^{ST}$), Universität Ulm, 89069 Ulm, Germany}

\begin{abstract}
Ultracold ground-state molecules can be formed from ultracold atoms via photoassociation followed by a spontaneous emission process. Typically, the molecular products are distributed over a range of final states. Here, we propose to use %carry out photoassociation in 
an optical cavity with high cooperativity to selectively enhance the population of a pre-determined final state by controlling the spontaneous emission. During this process, a photon will be emitted into the cavity mode. Detection of this photon heralds a single reaction.  We discuss the efficiency and the dynamics of cavity-assisted molecule formation in the frame of realistic parameters that can be achieved in current ultracold-atom setups. In particular, we consider the production of Rb$_2$ molecules in the $a^3\Sigma_u$ triplet ground state. Moreover, when working with more than two atoms in the cavity, collective enhancement effects in chemistry should be observable.
\end{abstract}

\maketitle

\section{Introduction}

In recent years, the control and manipulation of ultracold atomic samples has enabled studies of chemical reactions in the ultracold regime. Here, the internal and external quantum states of the collision partners can be very well controlled, allowing for precise studies of reactions and observations of possible quantum interference effects. Furthermore, it might be possible to gain absolute control over chemical reactions (reviews: \cite{Que12, Kre05}). %Similar ideas:  superchemistry \cite{Hei00}.
In ultracold chemistry, one important reaction type is photoassociation where laser light can fuse together colliding atoms into a well-defined excited molecular bound state \cite{Jon06,Ulm12}.  Typically, the excited molecule can decay spontaneously into a number of ro-vibrational levels in the electronic ground state.

Here, we propose a way to control the chemical reaction. For this, we combine concepts of cavity quantum electrodynamics (CQED) in an high-finesse optical cavity with ultracold molecule formation. We make use of the fact that strong confinement of the electromagnetic field modes around a molecule can control its spontaneous emission and thus its final quantum state. Our scheme is related to previous proposals for a molecular matter-wave amplifier in an optical cavity \cite{Sea04} or for coupling atoms to broadband photonic crystal waveguides \cite{Per17}.

Strong-light matter coupling with molecules is a recent hot topic, even outside the field of  ultracold temperatures. Applications range from quantum information processing to the modification of chemical reaction landscapes. In these contexts, coherent coupling of single dye molecules \cite{Wan17} and ensembles of polymers \cite{Sha15} to a cavity have recently been observed.

So far, high-finesse optical microcavities have been very successfully used for single cold atoms (for a review, see e.g., \cite{Reis15}). The strong confinement of the electromagnetic field in such cavities enables fast coherent transfer of atomic excitation into the cavity mode before the atom can decay by spontaneous emission. Instead, the cavity photon is emitted into a single external mode with high probabilty while the atom is prepared in a desired ground state. Important applications include deterministic single-photon sources \cite{Kuh02}.

Introducing micro-cavities technologies to the realm of ultracold chemistry will enable us to control and instantly detect single reactions with a high efficiency. Furthermore, it will allow for a detailed study of the  dynamics and statistics of reactions, in particular where collective effects come into play.

The implementation of cavity quantum electrodynamics concepts with molecules is even more challenging than for atoms: Due to the lack of closed electronic transitions in molecules and due to Franck-Condon factors which are in general small, smaller cavity mode volumes than for atoms are required to reach the regime of high cooperativity. Nevertheless, recent advances in cavity design and fabrication (e.g., fiber-based microcavities \cite{Hun10}) have enabled much smaller mode volumes, higher coupling strengths and better integrability compared to traditional cavities formed by bulky mirrors. Therefore, CQED technologies can now be applied to ultracold chemistry.

\begin{figure}
\begin{tabular}{cc}
\def\svgwidth{0.6\columnwidth}
(a)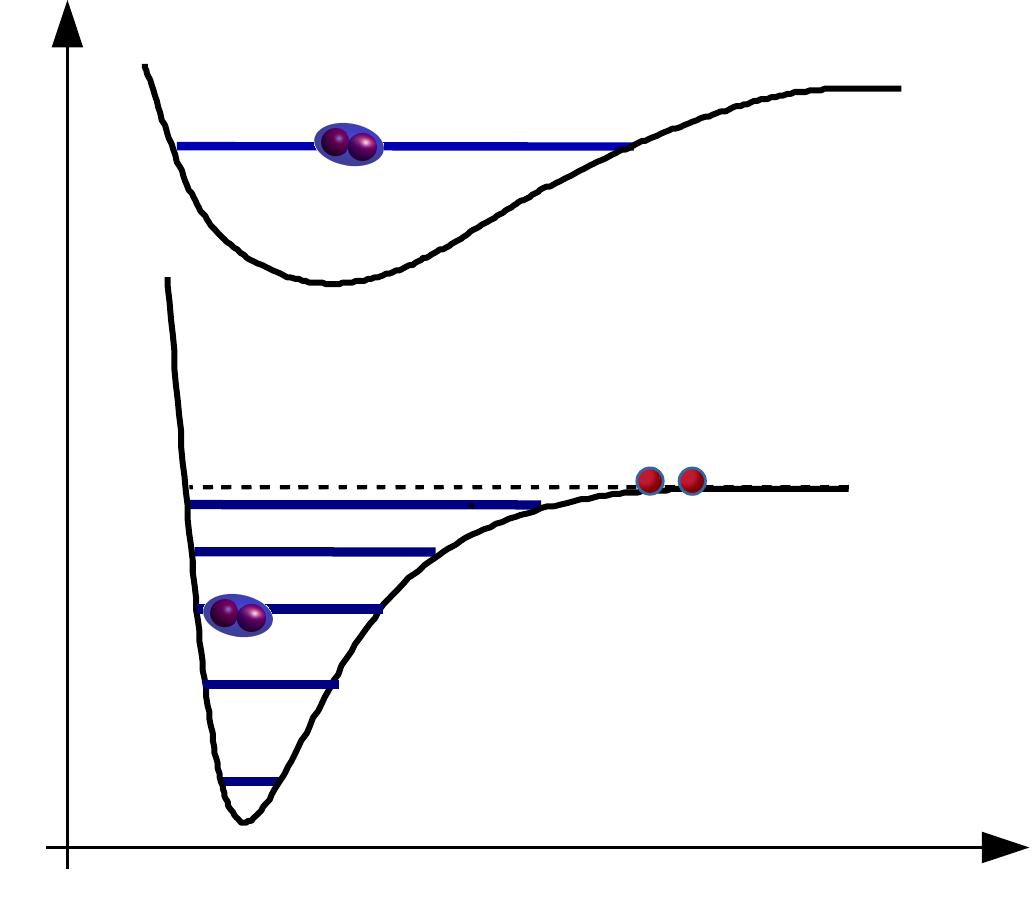&
\begin{minipage}{0.4\columnwidth}
\begin{flushleft}
\vspace{-4cm}
\def\svgwidth{2cm}
(b)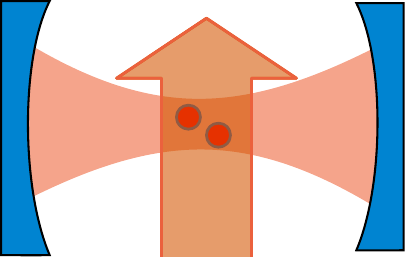
\def\svgwidth{2cm}
(c)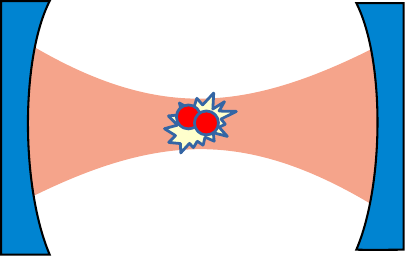
\def\svgwidth{2.5cm}
(d)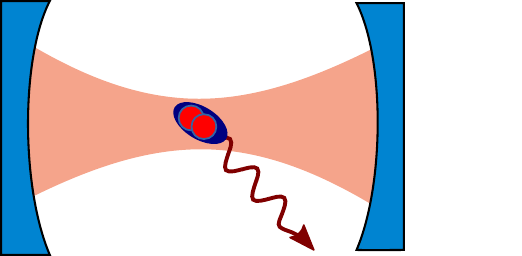
\end{flushleft}
\end{minipage}
\end{tabular}
\caption{(a) Cavity-controlled chemical reaction scheme for two atoms. Shown are the ground state and electronically excited potential energy curves and some bound states.
A photoassociation laser couples two unbound ground state atoms, denoted as $\ket i$, to a bound excited molecular state $\ket e$ with Rabi frequency $\Omega$, see also (b).  The cavity couples this state at rate $2g$ to a molecular ground state $\ket g$, see also (c). Alternatively, the molecule can decay to other states at rate $\Gamma$. (c) After an electronically excited molecule $\ket e$ has been formed, the excitation oscillates ($2g$) between the molecule and the cavity mode. (d) Either the excited molecule decays ($\Gamma$), or the cavity photon ($2\kappa$). In the latter case, the molecule is left in a predetermined ground state level $\ket g$ and the photon can be detected with high efficiency outside the cavity.}
\label{cavity_PA}
\end{figure}

After presenting the basic scheme for cavity-chemistry, we estimate realistic parameters for an experimental set-up with ultracold rubidium atoms. Afterwards, we simulate the reaction dynamics and efficiency for a square-shaped photoassociation pulse. Finally, we discuss collective effects when several molecules are produced.

\section{Single-molecule scheme}

%At the heart of the scheme lies the coupling of a molecular ground state level $|g\rangle$ to an electronically excited molecular state $|e\rangle$ by the resonator field , see Fig.~\ref{cavity_PA}(a). The coherent coupling strength is denoted by $g$, which quantifies the rate at which energy can oscillate between the molecule and the cavity field, see also Fig.~\ref{cavity_PA}(c). However, the excited state and the cavity field incoherently decay at rates $\Gamma$ and $\kappa$, respectively, see also Fig.~\ref{cavity_PA}(d).

We start out by presenting a somewhat simplified version of the cavity-controlled photoassociation scheme.  
We consider an unbound atom pair $\ket i$ which is trapped by an optical dipole trap in between the cavity mirrors. 
A bound state $\ket e$ of the electronically excited molecular potential (asymptotically, e.g., S+P) is excited from $\ket i$ by a photoassociation (PA) laser with Rabi frequency $\Omega$, see Fig.~\ref{cavity_PA}(a). The  laser illuminates the atoms from the open side of the cavity, see Fig.~\ref{cavity_PA}(b). Subsequently, the level $\ket e$ can spontaneously  decay to the molecular ground state manifold (S+S), typically within a few ns, as determined by the inverse linewidth $1/ \Gamma$ of the excited molecular level. In principle, there are many molecular ground state levels available for this decay. However, the cavity can be used to enhance the spontaneous decay into a particular level $\ket g$, and can therefore control the chemical reaction. For this, the cavity is tuned such that a cavity mode is resonant with the emitted photon. Under this condition, the molecule can undergo a transition from $\ket e$ to $\ket g$ while creating a photon in the cavity mode \cite{Sea04,Mor07,Lev08}. The rate of this coherent (reversible) energy exchange is denoted by $2g$, see Figs.~\ref{cavity_PA}(a,c). In addition, decoherence takes place: the excited state decays at rate $\Gamma$ and the cavity field at rate $\kappa$, see also Fig.~\ref{cavity_PA}(d). In the regime of high cooperativity $C\equiv g^2/(\kappa\Gamma)\gg1$, the electronic excitation can be transferred to the cavity mode faster than it can spontaneously decay into other free space modes. This means, the probability to get a ground-state molecule in a desired (ro-vibrational) quantum state can then be Purcell-enhanced. Since the photon decays from the cavity into a single external spatial mode, it can be detected with high efficiency, thus heralding every single reaction event by a photon click.

\subsection{Model Hamiltonian}     
  
   \begin{figure}
\def\svgwidth{0.8\columnwidth}
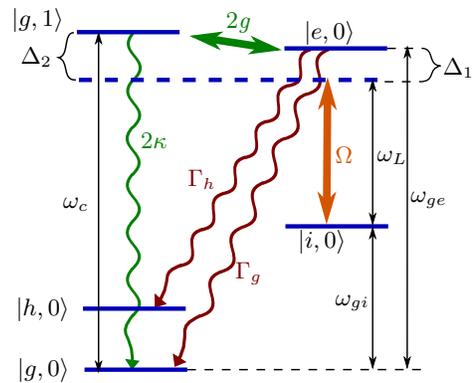
\caption{Quantum states which are involved in cavity-stimulated photoassociation with their detunings, couplings and decay rates, see text.}
\label{Cavity_PA_model}
\end{figure}
 
When driving the $\ket i \leftrightarrow \ket e$ transition for a {\em single} atom pair with the PA laser, at most one electronic or photonic excitation can be brought into the system at a time. We can then model the system effectively by five quantum states denoted by $\ket{m,n}$, where $m$ denotes the atomic/molecular quantum state and $n$ the cavity photon number, see Fig.~\ref{cavity_PA}. Furthermore, we assume tight confinement in the Lamb-Dicke regime ($\hbar\omega_\textrm{trap}\gg E_\textrm{recoil}$), and therefore, we need not consider the external motion of the particles.

   %of the asymptotic two-atom-state, two molecular ground states, one excited state and a cavity photon. 
   The couplings, detunings and decay rates of the five states are shown in Fig.~\ref{Cavity_PA_model}. The PA laser field (frequency $\omega_L$) couples the asymptotic two-atom-state $\ket{i,0}$ and the excited molecular state $\ket{e,0}$ with Rabi frequency $\Omega$ and detuning $\Delta_1=\omega_{ge}-\omega_L-\omega_{gi}$. The $\ket{e,0}$-state is coupled coherently at rate $2g$ to the state $\ket{g,1}$, which can decay to $\ket{g,0}$ at rate $2\kappa$. The two-photon detuning $\Delta_2=\omega_c-\omega_L-\omega_{gi}$ is the frequency mismatch between emitted photon and cavity frequency. The $\ket{e,0}$-state can also decay spontaneously at rates $\Gamma_g$ and $\Gamma_h$ to the states $\ket{g,0}$ and $\ket{h,0}$, respectively ($\Gamma=\Gamma_g+\Gamma_h$). Here, $\ket{g,0}$ represents the desired final state, while $\ket{h,0}$ represents all other possible states.
   
   The Hamiltonian of the coherently-coupled sub-system
   $\ket{i,0}, \ket{e,0}, \ket{g,1}$ 
   in a frame rotating with the laser frequency reads in the rotating-wave approximation
   \begin{align*}
   \hat H=&\hbar\Delta_1\ket{e,0}\langle e,0|+\hbar\Delta_2\ket{g,1}\langle g,1|\\
   &+\hbar \frac\Omega2 \ket{e,0}\langle i,0|+ \hbar g \ket{g,1}\langle e,0|+h.c.
   \label{Hamiltonian}
   \end{align*}
 To take into account the molecular and cavity decay processes, we solve the corresponding master equation in Lindblad form (see Appendix). The system is initialized in $\ket{i,0}$ at $t=0$.
 The goal is to transfer the population from this state % $\ket{i,0}$  
 as fast, efficiently and coherently as possible to the state  $\ket{g,0}$ of the ground state molecule. 
  We therefore define the efficiency $\eta$ of the scheme by the probability to produce a ground-state molecule $\ket{g,0}$ via cavity decay, 
 $$\eta=2\kappa \int_{t=0}^\infty\rho_{g1g1}(t) \textrm{d}t,$$
where $\rho_{g1g1}(t)$ is the population of the $\ket{g,1}$ state.

We can derive an analytical expression for $\eta$ in the weak-driving limit, ($\Omega\ll\Gamma,g^2/\kappa$), which, as we will later see, exhibits optimal achievable efficiency for time-independent control parameters. 
 In the weak-driving limit, the
coherent couplings of the laser $\Omega$ and the cavity $g$  slowly keep populating the spontanously decaying states $\ket{e,0}$  and $\ket{g,1} $. As a consequence, a quasi-steady superposition state $| D \rangle$ forms, which slowly decays in an exponential manner.
As derived in the appendix,
 \begin{equation}
\ket D\approx\ket{i,0}+\frac\Omega{2g^2+\Gamma (\kappa-i\Delta_2)}[ (i\kappa+\Delta_2) \ket{e,0} - g \ket{g,1}],
\label{eq:darkstate}
\end{equation}
when ignoring the slow exponential decay and assuming resonant drive, $\Delta_1=0$.
The decay takes place through the small populations in 
 $\ket{e,0}$ and $\ket{g,1}$ which decay via $\Gamma$ and $\kappa$, respectively.
% In the limit of weak driving ($\Omega\ll\Gamma,g^2/\kappa$), the system is most of the time in state $\rho_\textrm{wd}=\ket D\bra D$ (see appendix) where, for resonant driving ($\Delta_1=0$),
This yields the transfer efficiency,
%This almost dark state with small contributions from $\ket{e,0}$ and $\ket{g,1}$ slowly decays via $\Gamma$ and $\kappa$, respectively, yielding
 \begin{equation}
 \eta_\textrm{wd}=\frac{2\kappa\rho_{g1g1}}{2\kappa\rho_{g1g1}+\Gamma\rho_{e0e0}}=\frac{2C}{2C+1+(\Delta_2/\kappa)^2},
 \label{etamax}
 \end{equation}
where $\rho_{jj}$ are components of the desity matrix $\rho_\textrm{wd}=\ket D\bra D $. The efficiency $\eta_\textrm{wd}$ is maximal on two-photon resonance, $\Delta_2 =0$, which we consider from now on. One can also interpret $\eta_\textrm{wd}$ as the ratio of cavity-induced decay rate $\Gamma_\kappa=2g^2/\kappa$ of $\ket{e,0}$ (via green arrows in Fig.~\ref{Cavity_PA_model}) and the total decay rate $\Gamma_\kappa+\Gamma$ of $\ket{e,0}$ (for $\Delta_2 =0$). %This ratio is also often called $\beta=f/(f+1)$ in the literature, where $f=2C$ is the Purcell factor \cite{Kuh10}.
To obtain significant cavity-stimulated photoassociation, i.e., a large fraction in the $\ket{g,0}$ state, we therefore aim at $C \gg 1$. For example, a moderate cooperativity $C\gtrsim5$ will already result in an efficiency of $\eta_\textrm{wd}> 0.9$ for producing the chosen molecular quantum state.

As already mentioned the weak-driving limit gives optimal results in terms of transfer efficiency. This can be made plausible as follows. 
For vanishing $\Delta_2$ and $\kappa$, the states $\ket{i,0}, \ket{e,0}, \ket{g,1}$ form a $\Lambda$-system which exhibits a dark state $\ket D \propto 2g \ket{i,0} + \Omega \ket{g,1}$. This state is long-lived with no spontaneous decay via $\Gamma$, because $\ket{e,0}$ is not populated. If we now allow for a small but finite $\kappa\ll g^2/ \Gamma$ to populate the desired final state $\ket{g,0}$, the dark state $\ket D$ will become a little bit lossy due to population of $\ket{e,0}$ ($\propto \kappa^2$,  see eq.~(\ref{eq:darkstate})) and subsequent decay via $\Gamma$. These unwanted losses, however, are much smaller than the wanted decay flux from $| g,1\rangle \rightarrow | g,0\rangle $ which is proportional to $ \kappa$. In addition, in the weak driving limit, $\Omega \rightarrow 0$,  the dark state $| D \rangle $ mainly consists of state $|i,0\rangle$. Since this is precisely the initially prepared state the weak driving limit is optimal for efficient transfer from $|i, 0\rangle$ to the molcular state $|g, 0\rangle$.

%\tk{For sufficiently small $\Omega$ the state $\ket D$ mainly consists of state $\ket{i,0}$, which is the initially prepared state.}

% For vanishing $\Delta_2$ and $\kappa$, the states $|i,0\rangle, |e,0\rangle, |g,1\rangle$ form a $\lambda$-system which exhibits a dark state $ | D \rangle \propto 2g |i,0 \rangle + \Omega |g,1 \rangle  $. This  state is long-lived with no spontanous decay via $\Gamma$, because $|e,0 \rangle $ is not populated. 
%If we now allow for a non-zero but small-enough decay rate $\kappa << g^2/ \Gamma$, the dark state $| D \rangle$ will become a little bit lossy due to population of 
%  $|e,0 \rangle $ ($\propto \kappa^2$,  see eq.~(\ref{eq:darkstate})) and subsequent decay via $\Gamma$. 
%   
%
%

\subsection{Realistic experimental parameters}
Compared to a single atom, a molecule often exhibits a reduced dipole matrix element of the electronic transition and thus a reduced  coupling strength $g$ and cooperativity $C$. The reduction is to first approximation determined by the Franck-Condon-factor $f_\mathrm{FC}=\Gamma_g/\Gamma$ for the specific ro-vibrational transition in a molecule,
  $$g=g_\textrm{max}\sqrt{f_\mathrm{FC}}\quad\textrm{and}\quad C=C_\textrm{max}f_\mathrm{FC},$$
where the coupling strength and cooperativity for a closed two-level system ($f_\mathrm{FC}=1$) are given by
\begin{equation}
g_\textrm{max}=d_\text{el}\sqrt{\frac{\omega_{ge}}{2\hbar\epsilon_0V}}\quad\textrm{and}\quad C_\textrm{max}=\frac{g_\textrm{max}^2}{\kappa\Gamma},\label{gmax}
\end{equation}
respectively.
Here, $V$ is the volume of the cavity mode and $d_\text{el}$ the dipole moment of the electronic molecular $\ket g\leftrightarrow\ket e$ transition with frequency $\omega_{ge}$ (for a CQED review, we refer the reader to e.g.,~\cite{Reis15}). In a diatomic molecule, $d_\text{el}$ depends in general on the internuclear distance $R$, and the decay rate $\Gamma$ is about $2$ times larger as compared to an atomic excited state (due to a Dicke superradiance effect \cite{Dic54}).
   In a rubidium dimer (Rb$_2$), there are strong transitions between ro-vibrational states $\textrm{v}'$ of the shallow well of the $(1)^3\Pi_g$ potential (which asymptotically correlates to the atomic states $5\textrm{S}_{1/2} + 5\textrm{P}_{3/2}$) and the states $\textrm{v}''$ of the $a^3\Sigma_u$ potential ($5\textrm{S}_{1/2} + 5\textrm{S}_{1/2}$). % with a depth $\sim h\times7\,$THz \cite{Str10}. 
  % {\bf Für eine bessere Orientierung wäre es hier noch gut, zu wissen, welche Bindungsenergie  bzw. welche Vibrationsquantenzahlen diese Zustände haben. Kannst Du das raussuchen?}
 In particular, the transition $\textrm{v}'=8\rightarrow \textrm{v}''=0$ at a wavelength of $744\,$nm has the largest $f_\mathrm{FC,max}= 0.37$, see \cite{Bel11}. %However, it is somewhat close to the atomic resonance and therefore can be inconvenient. The second transition is at a wavelength of 993nm but is still sizeable,
 %  $f_\mathrm{FC} \gtrsim 0.1$. We already have experience  working with this transition.
  % transitions with moderate $f_\mathrm{FC} \gtrsim 0.1$ (transitions with $f_\mathrm{FC}> 0.3$ have been identified in Rb$_2$ \cite{Bel11}).
Apart from the transition dipole moment and Franck-Condon factor, the coupling strength $g$ depends also on the mode volume as $g\propto 1/\sqrt{V}$, thus $V$ should be minimized. For the fundamental TEM$_{00}$ mode of a Fabry-Perot resonator, $V=\pi w_0^2 L/4$, where $L$ is the cavity length and $w_0$ the mode waist.  
Similar as in \cite{Kam14}, we consider the dipole trap beams 
 to enter from the side without being clipped by the mirror substrates, which puts a lower limit on $L$. The mode waist $w_0$ is typically limited by the numerical aperture of the in-/outcoupling optics. 
 According to Eq.~\ref{gmax}, in order to maximise $C$, $\kappa$ should be minimized. However, as we will discuss in the next section a small $\kappa$ might lead to unacceptably long transfer times. Furthermore, the transmission through
 %To maximize the cooperativity $C$, the cavity field decay rate $\kappa$ has to be minimized, see Eq.~\ref{gmax}. However, $\kappa$ cannot be arbitrarily small because
 %the transmission through
   the mirror coatings should dominate over unavoidable absorption and scattering losses in the coatings. In table \ref{cavpar}, we give an example for realistic CQED parameters for the above mentioned molecular transition in Rb$_2$. For those parameters, an efficiency of $\eta_\textrm{wd}> 0.9$ could be achieved for vibrational transitions with $f_\mathrm{FC}\gtrsim 0.05$.

\begin{table}
\begin{tabular}{lcc}
\hline
Parameter & Symbol & Value\\
\hline
Cavity length & $L$ & $280\,\mu$m\\
Cavity mode waist & $w_0$ & $4.8\,\mu$m\\
Cavity finesse & $\mathcal{F}$ & $5\times 10^4$\\
Coupling strength for $f_\textrm{FC}=1$ & $g_\textrm{max}$ & $2\pi\times 80\,$MHz\\
Cavity field decay rate & $\kappa$ & $2\pi\times 5.4\,$MHz\\
Excited state decay rate & $\Gamma$ & $2\pi\times 12\,$MHz\\
Cooperativity for $f_\textrm{FC}=1$ & $C_\textrm{max}$ & $100$\\
Max.\ Franck-Condon factor & $f_\textrm{FC,max}$ & $0.37$ \cite{Bel11}\\
\hline
\end{tabular}
\caption{Example of a set of realistic CQED parameters. The parameters $g_\textrm{max}, \Gamma, C_\textrm{max}$ are derived for the transitions between vibrational states of the $(1)^3\Pi_g$ and $a^3\Sigma_u$ potentials in Rb$_2$. Here, $d_\textrm{el}\approx3\times10^{-29}\,$C$\cdot$m, see \cite{Per17,All12}.}
\label{cavpar}
\end{table}

\subsection{System dynamics}
Ideally, the transfer from the atom pair $|i, 0\rangle$ to the molecular state 
 $|g, 0\rangle$ should be efficient and fast. 
In the weak driving limit the transfer efficiency is nearly optimal but can be very slow as the
transfer time is $\propto \kappa^{-1}$. 
%{\bf JHD: Achtung, ist die folgende Formel richtig? Ich vermisse die Abhängigkeit/ Proportionalität zu $\kappa$.  Die Formel für die "kohärent" produzierten Moleküle sollte keine "1" im Nenner haben, denke ich. } 
In fact, it can be shown that the
transfer rate, which is the exponential decay rate of $\ket D$, is given by
%In the weak driving limit, we can reach $\eta_\mathrm{wd}$.
% However, it can be shown that the corresponding molecule formation rate
\begin{equation}
R_\mathrm{wd}=\frac{\Omega^2}{\Gamma(2C+1)}\underset{C\gg1}{=}\kappa \frac{\Omega^2}{2g^2},
\label{Rwd}
\end{equation}
%can become quite low
 if $\Delta_1=0$. 
 % This is essentially a slow pumping process where the initial state $\ket{i,0}$ is exponentially depumped on the timescale $1/R_\mathrm{wd}$. From an experimental point of view, 
 Clearly, too slow a transfer can be problematic for many reasons, e.g. when the transfer time is on the order of the particle lifetime in the trap. 

We therefore now consider the limit of strong driving where a very short $\pi$-pulse 
quasi instantaneously transfers the population from $\ket{i,0}$ to $\ket{e,0}$ at $t=0$.
 From there, it  partially decays into $\ket{g,0}$ via the cavity mode but it also partially decays via $\Gamma$ into $\ket{h,0}$. 
The  efficiency $\eta_\pi$ of the short-pulse scheme  is then reduced as compared to $\eta_\mathrm{wd}$ of the  weak-driving limit,
%In the simplest scheme, the population would be transferred quasi instantaneously from $\ket{i,0}$ to $\ket{e,0}$ at $t=0$, e.g. by using a short $\pi$-pulse.
% From there, it would partly decay into $\ket{g,0}$ via the cavity mode. 
 \begin{equation}
\eta_\pi=\frac{2\kappa}{2\kappa+\Gamma}\eta_\mathrm{wd},
\label{sdr}
 \end{equation}
where we set $\omega_c=\omega_{ge}$.
 %This is because all population is initially in $\ket{e,0}$ and can decay by spontaneaous emission before the population ratio between $\ket{e,0}$ and $\ket{g,1}$ relaxes to a fixed value, which is the case for weak driving.
Besides the decay, the dynamics exhibit a damped oscillation of the population  between $\ket{e,0}$ and $\ket{g,1}$. The oscillation can be understood as a beat of the two eigenstates $$\ket{B_\pm} \approx \frac1{\sqrt2}(\ket{e,0}\pm\ket{g,1})$$ of the coupled system,
which are populated at $t=0$ by the $\pi$-pulse as an equal superposition state. 
%This is because the $\pi$-pulse transfers all population at $t=0$ into an equal superposition of the  bright states 
%$$\ket{B_\pm} \approx \frac1{\sqrt2}(\ket{e,0}\pm\ket{g,1})$$ of the coupled system, i.e. the population oscillates between $\ket{e,0}$ and $\ket{g,1}$ and can first decay by spontaneaous emission before it decays via the cavity. In contrast, in the weak driving limit, the bright states are weakly populated and the system is mainly in the almost dark state $\ket D$.
%$$
%|D\rangle\approx\ket{i,0}+\frac\Omega{2g^2+\kappa\Gamma}( i \kappa \ket{e,0} - g \ket{g,1}), 
%$$
%which has only a small contribution from $\ket{e,0}$.
In order for $\eta_\pi$ to reach the weak-driving efficiency $\eta_\mathrm{wd}$, we need $\kappa\gg\Gamma$, according to eq.~(\ref{sdr}). However, for a large $\eta_\mathrm{wd}$ we need $\kappa \ll g^2/\Gamma  $. Both conditions can be  simultanously met only when $g \gg \Gamma$  \footnote{It turns out, that for a given $g$ the optimal  $\eta_\pi$ is reached for  $\kappa = g \gg \Gamma$}. According to our parameters in table~\ref{cavpar} and even when using a relatively large Franck-Condon factor of  $f_\mathrm{FC} > 0.3$, this condition is not easily reached. Assuming a more typical $f_\mathrm{FC}=0.1$, we obtain $\eta_\pi\approx 0.77$ compared to $\eta_\mathrm{wd}\approx 0.95$. 
Therefore, in general, the infinitly short excitation pulse does not seem to be ideal for the transfer. 
%However, in a realistic cold-atom experiment, it is hard to make $\kappa$ large enough while still maintaining a large $C$ and corresponding $\eta_\mathrm{wd}$, because of the above mentioned limitations on the mode volume and corresponding coupling strength $g$. For a given $g$, the maximum of $\eta_\pi$ is reached for $\kappa=g$. 
%For our example parameters, see table~\ref{cavpar}, assuming $f_\mathrm{FC}=0.1$, this would result in $\eta_\pi\approx 0.77$ compared to $\eta_\mathrm{wd}\approx 0.95$.

\begin{figure}
\def\svgwidth{\columnwidth}
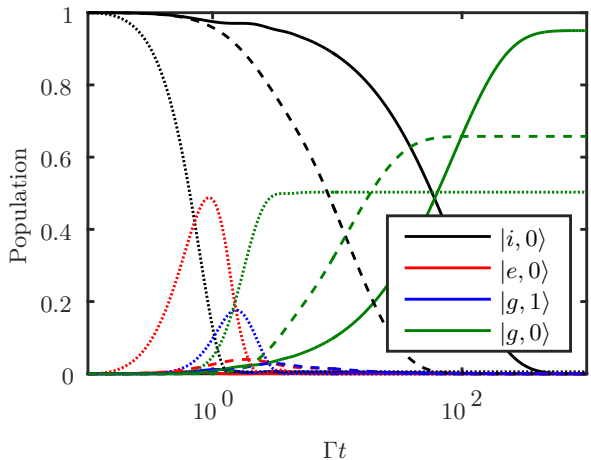
\caption{Single-molecule population dynamics of resonant cavity-stimulated photoassociation using a square PA laser pulse which is switched off when $99.9\%$ of the population from $\ket{i,0}$ has been transferred to $\ket{e,0}$. Shown are the cases $C=10$ (solid lines) and $C=1$ (dashed lines) for $\Omega=\kappa/2$, and $C=1, \Omega=3\kappa$ (dotted lines). For all cases, $\kappa=\Gamma$. If $\Omega$ is too high, the efficiency significantly decreases.}
\label{Cavity_PA_dynamics}
\end{figure}

\begin{figure}
\begin{flushleft}
(a)\\
\def\svgwidth{\columnwidth}
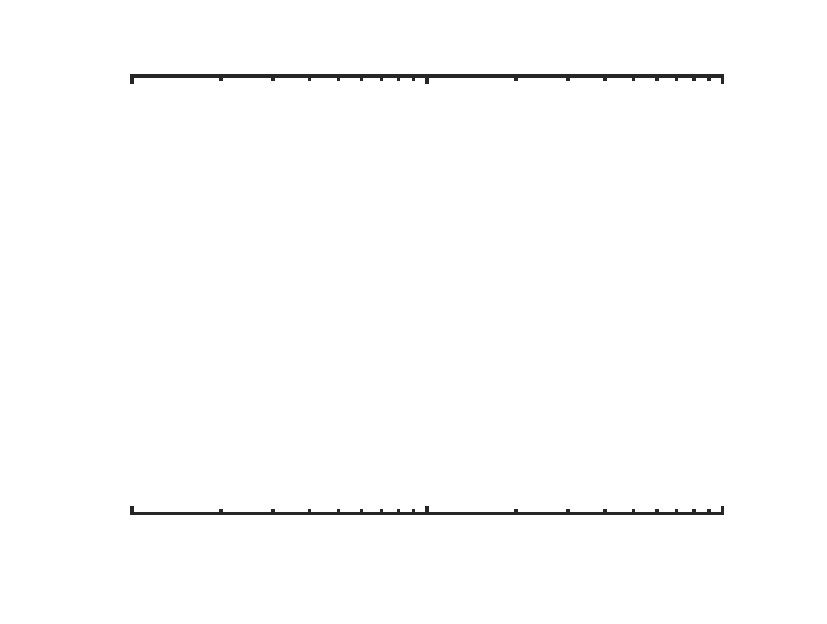
(b)\\
\def\svgwidth{\columnwidth}
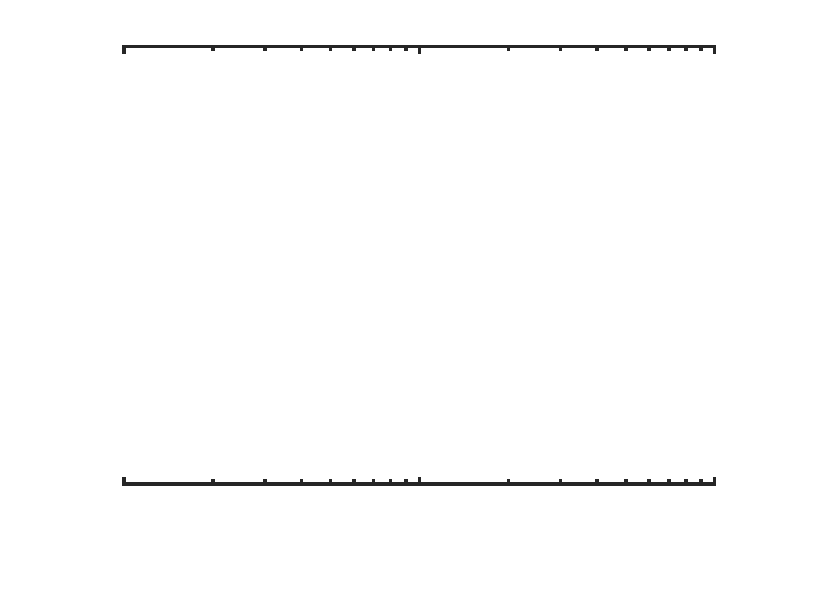
\end{flushleft}
\caption{(a) Inefficiency $1-\eta$ of resonant cavity-stimulated photoassociation using a square PA laser pulse. Here, we assume $g_\textrm{max}=2\pi\times 80\,$MHz and we chose $\varepsilon=0.1$. (b) Corresponding duration $t_p$ (solid lines) and Rabi frequency $\Omega$ (dashed lines) of the PA laser pulse.}
\label{Cavity_PA_Omega}
\end{figure}

Therefore, the question arises, how quickly the optical pumping can  be done without  substantial loss of efficiency compared to the weak-driving limit?
To answer this question, we investigate how the transfer efficiency to the ground state $\ket{g,0}$ can be optimized when employing a resonant square pulse of duration $t_p$ and Rabi frequency $\Omega$ \footnote{\protect{We note that there are more efficient schemes than using resonant square laser pulses to transfer electronic excitation into the cavity. Some of them are based on vacuum stimulated rapid adiabatic passages \cite{Hen00,Kuh10} and use specially designed laser pulses $\Omega(t)$ to transfer the system adiabatically from $\ket{i,0}$ to $\ket{g,0}$ without populating the excited state $\ket{e,0}$, i.e.\ by keeping it in a dark state. Those schemes are, however, beyond the scope of the present work.}}.

% for exciting the tranistion between $\ket{i,0}$  and $\ket{e,0}$.
% just as fast to $\ket{e,0}$ as it can decay to $\ket{g,0}$ via the cavity on a similar timescale. 

For this, we solve the time-dependent master equation of the five-level system numerically. In Fig.~\ref{Cavity_PA_dynamics}, some examples of the time-dependent populations are plotted. 
For $\Omega=\kappa/2$, we almost reach the weak-driving efficiencies $\eta_{\textrm{wd}}=\frac{20}{21}$ ($\eta_{\textrm{wd}}=\frac23$) for $C=10$ ($C=1$), respectively. Using $\Gamma=2\pi\times12\,$MHz, one reaches $95\%$ of the respective asymptotic value within $t\approx3\,\mu$s ($t\approx0.5\,\mu$s). For a higher Rabi frequency (here, $\Omega=3\kappa$), the transfer works faster but the efficiency is already significantly reduced. 

In order to carry out an optimization, we determine the highest Rabi frequency $\Omega$ and shortest pulse duration $t_p$ for which the {\it inefficiency} $1-\eta$ does not increase by more than a factor $1+\varepsilon$ compared to the weak-driving limit. In Fig.~\ref{Cavity_PA_Omega}, the inefficiency, pulse duration and the Rabi frequency are plotted for $\varepsilon=0.1$, assuming  $g_\textrm{max}=2\pi\times 80\,$MHz (see table \ref{cavpar}). We compare four different values of $f_\textrm{FC}$ and corresponding $g=g_\textrm{max}\sqrt{f_\textrm{FC}}$.
For larger $\kappa$ and smaller $f_\textrm{FC}$, shorter pulses with higher Rabi frequencies can be used, while the inefficiency increases (because $C$ decreases). For $\kappa/2\pi\lesssim 10$\,MHz, the Rabi frequency and the pulse duration roughly scale as $\Omega\approx\kappa$ and $t_p\propto f_\textrm{FC}/\kappa^2$. For larger values of $\kappa$, $\Omega$ diverges and $t_p$ vanishes (the condition for $t_p\rightarrow 0$ is derived in the appendix). For example, if an efficiency of $\eta=0.97$ is desired on a transition with $f_\textrm{FC}=0.1$, we require $\kappa=2\pi\times 3\,$MHz, see Fig.~\ref{Cavity_PA_Omega}(a). For this, we obtain $t_p=4\times10^{-5}\,$s and $\Omega=2\pi\times 3\,$MHz, see Fig.~\ref{Cavity_PA_Omega}(b).

%  as we will see in the next section.
 
%For $C\gtrsim3$, we find an almost linear dependency $\Omega\approx\kappa$. The corresponding pulse duration (after which $99.9\%$ of $\ket{i,0}$ is depopulated) is $t_p\approx10/R_\mathrm{wd}(\Omega,C)$, and can become quite long for large $C$ and small $\kappa$. Condition~(\ref{pulsecond}) is strongly violated in the $\kappa/\Gamma$-range shown.

%In contrast, for lower cooperativities $C\lesssim3$, we approach condition~(\ref{pulsecond}) with increasing $\kappa/\Gamma$. Therefore, the usable Rabi frequency $\Omega$ diverges, i.e.\ it can be made much larger than $\kappa$ without losing efficiency. Here, we are in the so-called ``fast cavity regime'' where $\kappa>g>\Gamma$. This regime with low $C$ 

\section{Collective effects}
\begin{figure}
\def\svgwidth{\columnwidth}
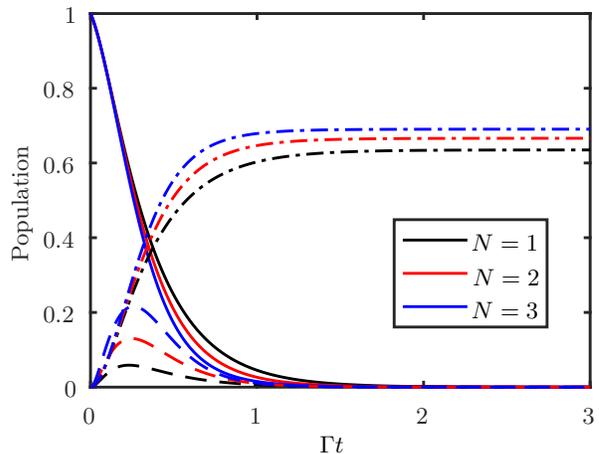
\caption{Collective dynamics: A sample of $N$ molecules is initially prepared in the $\ket{e,0}$ state (solid lines), from where it decays via the $\ket{g,1}$ state (dashed lines) into $\ket{g,0}$ (dashed-dotted lines). Both the decay rate and the final population in $\ket{g,0}$ increase with $N$. The parameters are: $C=1, \kappa=10\Gamma$.} 
\label{coll_dyn}
\end{figure}

So far, we have only considered the coupling of a single molecule to the cavity mode. However, in a typical experiment, there can be up to $N\approx 10^3$ atom pairs or molecules that couple simultaneously to the cavity mode (see Appendix). Coherence can build up among the molecules which modifies their spontaneous emission, thus they should not be treated as independent, giving rise to effects like Dicke-superradiance \cite{Dic54}. For example, if several of them are prepared in $\ket{e,0}$ at $t=0$, both the decay rate from this state and the probabilty to end up in the final state $\ket{g,0}$ are collectively increased, see Fig.~\ref{coll_dyn}. Here, the time-dependent master equation has been solved for $N=1\ldots3$ individual four-level molecules coupled to the same cavity mode. Our calculations show that for only up to three atom pairs enhancement effects are already present, but not very strong yet. We expect the enhancement to strongly increase with larger particle numbers. However, for the corresponding calculations other numerical methods  than the one used here would be required.

%These are first signs of the enhancement effect. For larger molecule numbers, other numerical methods than the one used here would be required. 
%The behaviour can be understood in terms of Raman lasing or superfluorescence, where a cavity field builds up that stimulates the $|e\rangle\leftrightarrow|g\rangle$ transition. %In Ref.~\cite{Sea04}, the effect is described in terms of coherently amplified scattering between different bosonic modes. 

The collective enhancement is most effective in the ``fast cavity regime'' where $\kappa>g>\Gamma$. Although  the transfer for single atom pairs  in this regime is plagued by
relatively low efficiencies $\eta$, the collective enhancement can strongly compensate for this.
 Furthermore, the collective enhancement is particularly suitable to enhance transitions with rather small Franck-Condon factors.

\section{Conclusions and outlook}
In summary, we have shown that ultracold molecule formation in a well-defined quantum state can be strongly enhanced by an optical cavity. We have estimated that preparation of a molecule in certain quantum states can be efficient ($>90\%$) for transitions with moderate Franck-Condon factors ($f_\textrm{FC}\gtrsim 0.05$) in medium high finesse cavities ($\mathcal F\approx5\times10^4$). In contrast to existing coherent two-photon schemes for photoassociation of ultracold molecules, our scheme can be regarded as a method to pump (cool) molecules into a desired (ground) state since the necessary dissipation is already build in via the cavity losses. Moreover, the photons dissipated into a single spatial mode can be detected with high efficiency. 

The scheme could potentially be extended to cascasded reactions, e.g., $A\rightarrow B^*\rightarrow C^*\rightarrow D$, where $B^*$ is, e.g., produced in a collision. Here, one could use the cavity to observe and control the spontaneous transition between $B^*\rightarrow C^*$. 

Finally, the coupling of several atom pairs or molecules to the cavity will give rise to interesting collective effects for reactions, which have not been discussed or demonstrated yet. The collective enhancement of the rate and efficiency of molecule formation would be the basis for ``superradiant chemistry''.

\section{Acknowledgments}
T.K.\ acknowledges a Marie Sk{\l}odowska-Curie postdoc fellowship by the European Commission (Standard EF, GA No.\ 747160) and a young researcher grant by the Baden-Württemberg Foundation and the Center for Integrated Quantum Science and Technology (IQ$^\textrm{ST}$).

\section{Appendix}
\subsection{Master equation}
In order to take into account the incoherent decay processes of the excited molecular state and the cavity photon, we use a master equation in Lindblad form,

\begin{align*}
 \frac{\textrm{d}\hat\rho}{\textrm{d}t} = &-\frac{i}{\hbar}[\hat H, \hat \rho]\\
 &+2\kappa\mathcal{D}[\hat a,\hat \rho]+ \Gamma_g\mathcal{D}[\ket g\bra e,\hat \rho]+ \Gamma_h\mathcal{D}[\ket h\bra e,\hat \rho],\\
 &\textrm{where }\mathcal{D}[\hat b,\hat \rho]=\hat b \hat\rho \hat b^\dagger-\frac12\hat b^\dagger\hat b\hat\rho-\frac12\hat\rho\hat b^\dagger\hat b.
 \end{align*}
 
 Here, $\hat \rho$ denotes the density operator and $\hat a$ is the annihilation operator of the cavity field.

\subsection{Efficiency in the weak-driving limit}
In the limit of weak driving ($\Omega\ll\Gamma,g^2/\kappa$), we can derive an analytical expression for the efficiency $\eta_\textrm{wd}$, Eq.~(\ref{etamax}). It is given by the ratio of the decay rates from $\ket{g,1}$ and $\ket{e,0}$ in ``quasi steady-state'', 
$$\eta_\textrm{wd}=\frac{2\kappa\rho_{g1g1}^\textrm{ss}}{2\kappa\rho_{g1g1}^\textrm{ss}+\Gamma\rho_{e0e0}^\textrm{ss}}=\left[1+\frac{\Gamma}{2\kappa}\frac{\rho_{e0e0}^\textrm{ss}}{\rho_{g1g1}^\textrm{ss}}\right]^{-1}.$$
Here, $\rho_{e0e0}^\textrm{ss}$ and $\rho_{g1g1}^\textrm{ss}$ are components of  the steady-state density matrix 
$\rho^\textrm{ss}$ which we calculate as follows. 
%To calculate the steady-state density matrix $\rho^\textrm{ss}$, 
We formally close the system in Fig.~\ref{Cavity_PA_model} by introducing an artificial ``repump'' rate $\zeta$ from the states $\ket{g,0}$ and $\ket{h,0}$ back to state $\ket{i,0}$,% see Fig.~\ref{Cavity_PA_mode_5lev_zeta}, 
which can be, in principle, arbitrarily slow ($\zeta\rightarrow 0$). It turns out, however, that the population ratio  $\rho_{e0e0}^\textrm{ss}/\rho_{g1g1}^\textrm{ss}$ is independent of $\zeta$,  turning the system effectively into a  three-level system, since the populations of the levels $\ket{g,0}$ and $\ket{h,0}$ are not relevant. %{\bf Warum?	Das kann ich nicht nachvollziehen.}
The solution for the density matrix  $\rho^\textrm{ss}=\ket{D^\textrm{ss}}\bra{D^\textrm{ss}}$ in the weak-driving regime is, to first order in  $\Omega$,
$$
\ket{D^\textrm{ss}}\approx\ket{i,0}+ \Omega \frac{(i\kappa+\Delta_2) \ket{e,0} - g \ket{g,1}}{2g^2+(\Gamma-2i\Delta_1)(\kappa-i\Delta_2)},
$$

If the system is open, i.e. $\zeta=0$, 
this state slowly decays exponentially, which can be described by 
$$\ket D =\exp\left\{- \frac{R_\textrm{wd}}2 t\right\}\ket{D_\textrm{ss}},$$
where $R_\textrm{wd}$ is given, for $\Delta_1=\Delta_2=0$, by eqn.~(\ref{Rwd}).
%Here $\gamma$ can be determined from the populations in  $\ket{g,1}$ and  $\ket{e,0}$ and their decay rates $\kappa$ and $\Gamma$, respectively.

\subsection{Condition for $\pi$-pulses}
For a chosen $\varepsilon$, infinitely short $\pi$-pulses (as discussed before) are efficient enough if 
 \begin{equation*}
(1-\eta_\pi)\leq(1+\varepsilon)(1-\eta_\textrm{wd}).
\end{equation*}
Using eqns.~(\ref{etamax}) and (\ref{sdr}), this is equivalent to the following condition
 \begin{equation*}
\frac\kappa\Gamma \geq \sqrt{\frac{g^2}{\varepsilon\Gamma^2}+\frac1{16}}-\frac14,
\label{pulsecond}
\end{equation*}
which can be approximated by $\kappa\gtrsim g/\sqrt{\varepsilon}$ for $g\gg\Gamma\sqrt{\varepsilon}$.

%this almost dark state with small contributions from $\ket{e,0}$ and $\ket{g,1}$ slowly decays into $\ket{h,0}$ and $\ket{g,0}$, respectively, yielding Eq.~(\ref{etamax}).
% 
%For arbitrarily slow driving ($\Omega\rightarrow 0$), we can derive the analytical expression for the efficiency $\eta_\textrm{wd}$. It is given by the ratio of the decay rates from $\ket{g,1}$ and $\ket{e,0}$ in ``quasi steady-state'',
% $$\eta_\textrm{wd}=\frac{2\kappa\rho_{g1g1}^\textrm{ss}}{2\kappa\rho_{g1g1}^\textrm{ss}+\Gamma\rho_{e0e0}^\textrm{ss}}=\left[1+\frac{\Gamma}{2\kappa}\frac{\rho_{e0e0}}{\rho_{g1g1}}\right]^{-1}.$$

%\begin{figure}
%\def\svgwidth{0.8\columnwidth}
%\input{figures/Cavity_PA_states2.pdf_tex}
%\caption{An artificial repump rate $\zeta$ is introduced into the five-level system of Fig.~\ref{Cavity_PA_model} to calculate steady-state populations analytically.{\bf Die Defintionen der Delta sind anders als in Fig. 2. }}
%\label{Cavity_PA_mode_5lev_zeta}
%\end{figure}

\subsection{Number of atom pairs}
The estimated number of pairs $N\approx 10^3$ in the optical cavity for typical experimental conditions  corresponds to a thermal cloud of approx.\ $10^6$ atoms with diameter $2\sigma\approx40\,\mu$m in a three-dimensional cubic optical lattice with laser wavelength $\lambda_\textrm{latt}\approx1\,\mu$m, which overlaps with the fundamental cavity mode with waist $w_0\approx 5\,\mu$m.

%\bibliography{papers_TK,papers_JHD,papers_others}

\end{document}